# From an obliquely falling rod in a viscous fluid to the motion of suspended magnetic bead chains that are driven by a gradient magnetic field and that make an arbitrary angle with the magnetic force vector: A Stokes flow study

Robert J. Deissler[1*], Rose Al Helo[1], Robert Brown[1]

[1]Department of Physics, Case Western Reserve University, Cleveland, Ohio, United States of America

[*]Corresponding author

E-mail: rjd42@case.edu

# Abstract

In view of the growing role of magnetic particles under magnetic field influence in medical and other applications, and perforce the bead chaining, it is important to understand more generally the chain dynamics. As is well known, in the presence of a magnetic field, magnetic beads tend to form chains that are aligned with the magnetic field vector. In addition, if there is a magnetic field gradient, there will be a magnetic force acting on this chain. The main goal of the present research is to study the motion of a magnetic bead chain that makes an arbitrary angle with the magnetic force vector in the Stokes flow limit, that is, in the limit of zero Reynolds number. We used the public-domain computer program HYDRO++ to calculate the mobility matrix, which relates the magnetic force acting on the chain to the velocity of the chain, for a chain of $N$ beads making an arbitrary angle with the magnetic force vector. Because of the presence of off-diagonal elements of the mobility matrix, as the chain is drawn in the direction of the magnetic force, it is also deflected to the side. We derived analytic solutions for this motion. Also, for bead chains moving in directions both parallel and perpendicular to their lengths, we fit three-parameter functions to solutions from HYDRO++. We found the fits to be excellent. Combining these results with the analytic solutions, we obtained expressions for the velocity components for the bead chains that provide excellent fits to HYDRO++ solutions for arbitrary angles. Finally, we apply the methodology used for the bead chain studies to the study of an obliquely falling rod in a viscous fluid and derive analytic solutions for the velocity components of the obliquely falling rod.

# Introduction

The study of magnetophoresis, or the motion of magnetic particles suspended in a fluid under the influence of a non-uniform magnetic field, is of increasing importance in many applications. Recent reviews [1–4] draw attention to the use of magnetic particles, particularly in the manipulation of cells in microfluidic devices, including their capture, and in targeted drug delivery. A focus on the magnetic particles themselves is necessary because of mutual interactions arising from their magnetic dipole moments, which are induced by the magnetic field. Of special interest is the emergence of chains and hence a connection to the motion of rods in a viscous fluid, in view of the force on magnetic moments arising from the gradient magnetic field.

Yang et al. [5] have presented experimental work measuring the drag coefficients of micron-sized magnetic beads and chains of such beads. These measurements were compared to various analytical predictions – ellipsoid, cylinder, and bead models – and also to the HYDRO++ numerical program [6–8]. This program is most successful in the description of the translational drag coefficients for a chain moving in a direction parallel to it length and for a chain moving in a direction perpendicular to its length. There was good agreement over a range of bead numbers making up a chain.

The work of [5] is relevant to a variety of applications including propulsion of organisms and robots in granular media [9], non-Newtonian fluid flow in metallurgy [10], rheological properties of hollow magnetite chains [11], sedimentation of Brownian chains [12], Stokes flow of protein crystals toward a magnet source [13], as well as vertical and lateral drag on cellular membranes [14]. These references include aspects and experimental implications of the modeling of chains addressed in [5] and earlier work on rigid bead [15,16], ellipsoid [17–21], and cylinder [18,20,22–25] models.

In view of the growing role of magnetic particles under magnetic field influence in medical and other applications, and perforce the bead chaining, it is important to understand more generally the chain dynamics. From the work on obliquely falling rods in a gravitational field in a viscous fluid [26–28], while the gravitational force and the magnetic force are very different, a richer motional picture for the magnetic bead chains is expected. We see lateral motion for falling rods [26–28] and sedimentation of colloidal chains [12], and thus anticipate analogous results for magnetic particle chains oriented at arbitrary angles to the magnetic force.

Consider a suspension of superparamagnetic beads in a fluid. In the presence of a magnetic field, the field induces a magnetic dipole moment in the beads causing them to attract one another and to form chains [5,29,30]. The chains tend to align themselves along the magnetic field vector $\mathbf{B}$. If there is also a nonzero gradient in the field, there will be a magnetic force,

$$\mathbf{F}_N^{mag} = N\mathbf{F}_1^{mag} = N(\mathbf{m}\cdot\nabla)\mathbf{B}, \tag{1}$$

acting on a chain of $N$ beads causing the chains to move in the fluid, where $\mathbf{m}$ is the induced magnetic dipole moment and $\mathbf{F}_1^{mag}$ is the magnetic force on a single bead. Because of motion in the fluid, there will be a frictional drag force acting on the chains. It is assumed that variables are such that the Reynolds number $\mathrm{Re} = \rho U \ell / \eta \ll 1$, where $\rho$ is the density of the fluid, $U$ is the speed of the beads relative to the fluid, $\ell$ is a characteristic length, and $\eta$ is the dynamic viscosity of the fluid. This type of flow, referred to as Stokes flow or creeping flow (with $\mathrm{Re} \to 0$), assumes that viscous forces dominate and that inertial forces may be neglected.

The study of [5] compared experimental results to various modeling approximations for the translational drag coefficients for magnetic bead chains in Stokes flow, including values obtained from the HYDRO++ numerical program [6–8]. Two cases were considered: 1) $(\mathbf{m}\cdot\nabla)\mathbf{B} \parallel \mathbf{B}$, causing the chains to move in a direction parallel to their length and 2) $(\mathbf{m}\cdot\nabla)\mathbf{B} \perp \mathbf{B}$, causing the chains to move in a direction perpendicular to their length.

The drag force acting on a chain of $N$ beads is given by [5]

$$\mathbf{F}_N^{drag} = -\zeta_N \mathbf{U}_N, \tag{2}$$

where $\mathbf{U}_N$ is the translational velocity of the chain and $\zeta_N$ is the drag coefficient. The magnetic force given by Eq (1) is balanced by the drag force giving

$$N\mathbf{F}_1^{mag} = \zeta_N \mathbf{U}_N. \tag{3}$$

Dividing the magnitude of Eq (3) for a general value of $N$ by the magnitude of Eq (3) for $N = 1$ gives

$$\frac{\zeta_N}{\zeta_1} = \frac{NU_1}{U_N}. \tag{4}$$

As pointed out in [5], values of the drag coefficients obtained from the public-domain computer program HYDRO++ [6–8] compare best to experiment.

For the two cases considered in [5], where the magnetic force vector is either parallel or perpendicular to the length of the chain, the velocity of the chain is in the same direction as the magnetic force. This may be understood in terms of the chain being symmetric about the magnetic force vector. An interesting question is what happens if the chain makes an angle with the magnetic force vector, other than $0°$ or $90°$?

As is well known in the case of the force due to gravity, consider an obliquely falling rod through a viscous fluid [26–28]. As a result of the interaction of the fluid with the rod, there is not only the vertical component of velocity of the rod in the direction of the gravitational field, but also a sideways component of velocity. This sideways motion is related to the fact that the drag coefficient of a vertically falling rod is less than that of a horizontally falling rod and may be understood by resolving the drag force on an obliquely falling rod into a component along the rod and a component transverse to the rod. Since the drag coefficient for motion along the rod is less than that for motion transverse to the rod, the rod will be deflected to the side. Since the drag coefficient for motion along a bead chain is less than that for motion transverse to the bead chain, we would expect a similar sideways component of velocity for the magnetic bead chains studied here.

In this paper we consider the case where $(\mathbf{m}\cdot\nabla)\mathbf{B}$ is neither parallel nor perpendicular to $\mathbf{B}$, implying that the magnetic force is neither parallel nor perpendicular to the length of the chain. Since the drag coefficient for a bead chain moving perpendicular to its length is greater than the drag coefficient for a bead chain moving parallel to its length, we find, as expected, that there is a sideways motion for the chain, like what occurs for the obliquely falling rod. As the chain is drawn in the direction of the magnetic force, the chain is also deflected to the side due the interaction of the chain with the fluid, causing a component of velocity perpendicular to the magnetic force vector. We derive an analytical expression for the angular dependence of this chain velocity.

We note that previous gravitational analyses [26–28] of an obliquely falling rod do not calculate the horizontal and vertical components of the velocity, but calculate only the deflection angle, and then only for the special case of a long thin rod, where the velocity of a vertically falling rod is twice that of a horizontally falling rod. We apply the methodology developed here for the bead chains, to the case of the obliquely falling rod, giving a general expression for the horizontal and vertical components of velocity in terms of the velocity of a vertically falling rod, the velocity of a horizontally falling rod, and the angle of the rod to the vertical.

## Methods

We expand the use of the public-domain computer program HYDRO++ [6–8] to solve for the more general velocity of the bead chain. For given positions of the magnetic beads, this program solves for the diffusion matrix. For Stokes flow, the diffusion matrix $\mathbf{D}$ is proportional to the mobility matrix $\mathbf{\mu}$ through the generalized Einstein equation as [31]

$$\mathbf{D} = k_B T \mathbf{\mu}, \qquad (5)$$

where $k_B$ is Boltzmann's constant and $T$ is the temperature. The mobility matrix relates the external force $\mathbf{F}$, which in our case is the magnetic force given by Eq (1), to the velocity of the chain $\mathbf{U}$ as [8,31]

$$\mathbf{U} = \mathbf{\mu}\cdot\mathbf{F}. \qquad (6)$$

In general, $\mathbf{D}$ and $\mathbf{\mu}$ are 6 x 6 matrices containing both translational and rotational components. Here, we are interested in only the translational components, although when running HYDRO++, the full diffusion matrix must be chosen. Also, we used ICASE=20 (third-order hydrodynamic interaction) when running HYDRO++, which we found to be the most accurate for the problem considered here. Note that it was not necessary to calculate the velocity of the bead chain by solving the differential equations for fluid flow, that is, the Stokes equation, $\nabla p = \eta \nabla^2 \mathbf{v}$ [26], where $p$ is the pressure, $\eta$ is the dynamic viscosity of the fluid, and $\mathbf{v}$ is the fluid velocity, with appropriate boundary conditions at the bead chain surface. It was only necessary to calculate the diffusion matrix from HYDRO++. The mobility matrix is then calculated from Eq (5), and the velocity of the chain from Eq (6). It should also be noted that the chain is assumed to be far from any boundaries, such as walls, to eliminate interactions between the chain and such boundaries.

The inverse of the mobility matrix is defined as the friction matrix $\boldsymbol{\zeta} = \boldsymbol{\mu}^{-1}$, which relates the velocity of the chain to the net force of the chain on the fluid as [8,31,32]

$$\mathbf{F} = \boldsymbol{\zeta} \cdot \mathbf{U}. \tag{7}$$

As noted in the Introduction, when the chain is either parallel or perpendicular to the magnetic force $\mathbf{F}^{mag}$, the velocity of the chain $\mathbf{U}$ is in the same direction as $\mathbf{F}^{mag}$. This is reflected in the fact that both the mobility matrix $\boldsymbol{\mu}$ in Eq (6) and the friction matrix $\boldsymbol{\zeta}$ in Eq (7) are diagonal matrices. This is confirmed for these cases by the fact that the diffusion matrix $\mathbf{D}$ in Eq (5) turns out to be diagonal when running HYDRO++. The components of the friction matrix for this case are equal to the drag coefficients referred to in Eqs (2), (3), and (4).

Also, as noted in the Introduction, if the chain makes an angle other than 0° or 90° with the magnetic force $\mathbf{F}^{mag}$, and as the chain is drawn in the direction of $\mathbf{F}^{mag}$, the interaction of the fluid with the chain causes the chain to be deflected to the side resulting in a sidewise component of velocity. Therefore, the velocity $\mathbf{U}$ is no longer in the direction of $\mathbf{F}^{mag}$. Referring to Eq (6), this would imply that the mobility matrix $\boldsymbol{\mu}$ is no longer diagonal, the off-diagonal elements being responsible for the sideways component of $\mathbf{U}$. As expected, we find that $\boldsymbol{\mu}$ has off-diagonal components when running HYDRO++ for this case.

## Results and discussion

### Fitting three-parameter functions to HYDRO++ calculations

To calculate the drag coefficients, the diffusion matrix is first calculated from HYDRO++, taking the beads to lie either on the *x*-axis or on the *y*-axis. Then, as noted in the previous section, the diffusion matrix and mobility matrix will be diagonal. Taking the magnetic force to be in the *x*-direction, we have from Eq (6)

$$\begin{pmatrix} U_{N,x} \\ U_{N,y} \end{pmatrix} = \begin{pmatrix} \mu_{N,xx} & 0 \\ 0 & \mu_{N,yy} \end{pmatrix} \begin{pmatrix} F_N^{mag} \\ 0 \end{pmatrix}. \tag{8}$$

Taking the chain to lie on the *x*-axis, we have $\zeta_N^{\parallel} = \mu_{N,xx}^{-1}$; and taking the chain to lie on the *y*-axis, we have $\zeta_N^{\perp} = \mu_{N,xx}^{-1}$. Alternatively, one can take the chain to lie on the *x*-axis, and instead change the direction of the magnetic force. Then, if the magnetic force is in the *x*-direction, we have $\zeta_N^{\parallel} = \mu_{N,xx}^{-1}$; and if the magnetic force is in the *y*-direction, we have $\zeta_N^{\perp} = \mu_{N,yy}^{-1}$. The drag coefficients are normalized to those for a single bead, by calculating $\zeta_1$ from HYDRO++ for a single bead and dividing it into the drag coefficients for a chain of $N$ beads.

Approximations for the drag coefficients, normalized by those for a single bead, for motion parallel and perpendicular to the length of the chains are given by the formulas [5,25]

$$\frac{\zeta_N^{\parallel}}{\zeta_1} = \frac{2}{3}\frac{N}{\ln N + \gamma_N^{\parallel}} \tag{9}$$

and

$$\frac{\zeta_N^{\perp}}{\zeta_1} = \frac{4}{3}\frac{N}{\ln N + \gamma_N^{\perp}}, \tag{10}$$

where $\gamma_N^{\parallel}$ and $\gamma_N^{\perp}$ are correction factors.

Reference [5] gives values for $\zeta_N^{\parallel}/\zeta_1$ and $\zeta_N^{\perp}/\zeta_1$ obtained from HYDRO++ which are provided in their table for values of $N$ from 2 through 30. In this paper we use HYDRO++ to calculate $\zeta_N^{\parallel}/\zeta_1$ and $\zeta_N^{\perp}/\zeta_1$ for values of $N$ from 2 through 50 and fit Eqs (9) and (10) to these values.

To this end, the correction factors are expressed in the references [5,25] by

$$\gamma_N^{\parallel} = b_0^{\parallel} + \frac{b_1^{\parallel}}{N} + \frac{b_2^{\parallel}}{N^2} \tag{11}$$

and

$$\gamma_N^{\perp} = b_0^{\perp} + \frac{b_1^{\perp}}{N} + \frac{b_2^{\perp}}{N^2}. \tag{12}$$

We vary the coefficients $b_0^{\parallel}$, $b_1^{\parallel}$, and $b_2^{\parallel}$, to give the best least squares fit of Eqs (9) and (11); and vary the coefficients $b_0^{\perp}$, $b_1^{\perp}$, and $b_2^{\perp}$ to give the best least squares fit of Eqs (10) and (12), all in comparison with the data from HYDRO++. A MATLAB code was written to perform the least squares fit. To obtain initial guesses for the coefficients, values for a cylinder of length $2rN$ and diameter $2r$ are used [5,25]: $(b_0^{\parallel}, b_1^{\parallel}, b_2^{\parallel})_0 = (-0.207, 0.980, -0.133)$ and $(b_0^{\perp}, b_1^{\perp}, b_2^{\perp})_0 = (0.839, 0.185, 0.233)$. Using these values as initial values for the iterative process, we obtain from our MATLAB code $(b_0^{\parallel}, b_1^{\parallel}, b_2^{\parallel}) = (-0.0689, 0.9743, -0.1041)$ and $(b_0^{\perp}, b_1^{\perp}, b_2^{\perp}) = (1.0333, 0.2506, 0.2780)$. So, we obtain the following equations for $\gamma$:

$$\gamma_N^{\parallel} = -0.0689 + 0.9743/N - 0.1041/N^2 \tag{13}$$

and

$$\gamma_N^{\perp} = 1.0333 + 0.2506/N + 0.2780/N^2. \tag{14}$$

Fig 1 shows a plot of the values of the least squares fit from Eqs (9), (10), (13), and (14), and from the values of HYDRO++ for the normalized drag coefficients for both motion parallel and perpendicular to the length of the chain. The agreement is seen to be excellent.

**Fig 1. Normalized drag coefficients.** Plot of the normalized drag coefficients for a chain of $N$ beads for motion both parallel and perpendicular to the length of the chain from the numerical solution using HYDRO++ and from Eq. (9), (10), (13), and (14), which resulted from a least squares fit to the numerical solution of HYDRO++. Agreement is seen to be excellent.

From Eq (4) we may also obtain values for the velocity of a chain normalized to the velocity of a single bead as

$$\frac{U_N}{U_1} = \frac{N\zeta_1}{\zeta_N}. \qquad (15)$$

Using the values of the normalized drag coefficients from Eqs (9), (10), (13), and (14), the normalized velocity of a chain with $N$ beads for both motion parallel and perpendicular to the length of the chain is shown in Fig 2.

**Fig 2. Normalized translational velocity.** Plot of the normalized velocity for a chain of $N$ beads for motion both parallel and perpendicular to the length of the chain from the numerical solution using HYDRO++ and from Eq. (9), (10), (13), (14), and (15) which resulted from a least squares fit to the numerical solution of HYDRO++.

## The velocity of chains for arbitrary angles

As noted in the Introduction, when the chain makes an angle -- other than $0°$ or $90°$ -- with the magnetic force vector, qualitatively new behavior occurs. As the magnetic force $\mathbf{F}^{mag}$ draws the chain toward the right, there is a drag force $\mathbf{F}^{drag}$ of the fluid on the chain that is equal and opposite to the magnetic force, as illustrated in Fig 3a. In addition, the interaction of the fluid with the chain causes the chain to be deflected to the side, producing an additional velocity component in the $+y$ direction, as illustrated in Fig 3b.

**Fig 3. Chain at an angle.** a) As the magnetic force draws the chain to the right in the $+x$ direction, the interaction of the fluid with the chain produces a drag force in the $-x$ direction. b) In addition, the interaction of the fluid with the chain causes the chain to be deflected to the side, producing an additional velocity component in the $+y$ direction, which is related to the off-diagonal elements of the mobility matrix.

Running HYDRO++ for a given value of $N$ and for a range of angles from $0°$ to $90°$ gives the diffusion matrix, which is now found to be non-diagonal and symmetric, and thus the mobility matrix $\boldsymbol{\mu}$ from Eq (5). Then from Eq (6) we obtain the velocity $\mathbf{U}$ of the chain, including the bead number $N$, as

$$\mathbf{U}_N = \boldsymbol{\mu}_N \cdot \mathbf{F}_N^{mag}. \qquad (16)$$

Taking $\mathbf{F}_N^{mag}$ to be in the *x* direction, and taking the chain to lie in the *x-y* plane, we find from Eq. (16)

$$\begin{pmatrix} U_{N,x} \\ U_{N,y} \end{pmatrix} = \begin{pmatrix} \mu_{N,xx} & \mu_{N,xy} \\ \mu_{N,xy} & \mu_{N,yy} \end{pmatrix} \begin{pmatrix} F_N^{mag} \\ 0 \end{pmatrix}, \quad (17)$$

where we have used the fact that $\mathbf{\mu}$ is symmetric. Therefore,

$$\begin{pmatrix} U_{N,x} \\ U_{N,y} \end{pmatrix} = \begin{pmatrix} \mu_{N,xx} F_N^{mag} \\ \mu_{N,xy} F_N^{mag} \end{pmatrix}. \quad (18)$$

As expected, we see that $\mu_{xy}$, the off-diagonal component of $\mathbf{\mu}$, is responsible for $U_y$, the sideways component of **U**. For a single bead we have

$$U_1 = \mu_1 F_1^{mag}. \quad (19)$$

Dividing Eq (18) by Eq (19), and introducing the notation that a tilde denotes a variable normalized to that of a single bead, we obtain

$$\begin{pmatrix} \tilde{U}_{N,x} \\ \tilde{U}_{N,y} \end{pmatrix} = \begin{pmatrix} \tilde{\mu}_{N,xx} \tilde{F}_N^{mag} \\ \tilde{\mu}_{N,xy} \tilde{F}_N^{mag} \end{pmatrix} = N \begin{pmatrix} \tilde{\mu}_{N,xx} \\ \tilde{\mu}_{N,xy} \end{pmatrix}, \quad (20)$$

where we have used, from Eq (1),

$$\tilde{F}_N^{mag} = \frac{F_N^{mag}}{F_1^{mag}} = N. \quad (21)$$

Fig 4 shows plots of $\tilde{U}_{N,x}(\theta)$ and $\tilde{U}_{N,y}(\theta)$ for $N=8$ from the solution of HYDRO++, as well as plots of cosine and sine functions. As seen in the figures, the cosine and sine functions fit precisely with the HYDRO++ solutions. This result can be understood by a straightforward proof as outlined in the next section.

**Fig 4. Normalized velocities in the *x* and *y* directions for a chain of 8 beads for a range of angles.** a) *x* component of the velocity. b) *y* component of the velocity. It is seen that the cosine and sine functions fit precisely with the HYDRO++ solutions. The coefficients *A*, *B*, and *C* are determined from $A = [\tilde{U}_x(0°) + \tilde{U}_x(90°)]/2 = 2.76758$, $B = [\tilde{U}_x(0°) - \tilde{U}_x(90°)]/2 = 0.42757$, and $C = \tilde{U}_y(45°) =$ 0.42701, respectively.

## Analytical solution for angular dependence of the velocity

Assume that the bead chain makes an angle of $\theta$ with the magnetic force. Noting the mobility matrix is diagonal for a chain aligned with the *x*-axis, we first write Eq (16) in a rotated coordinate system in which the chain is aligned with the *x'*-axis as illustrated in Fig 5. Noting that $\mathbf{F}^{mag}$ is in the *x* direction, and that the chain lies in the *x-y* plane, the velocity will lie in the *x-y* plane. Therefore, we have

$$\begin{pmatrix} \tilde{U}_{x'} \\ \tilde{U}_{y'} \end{pmatrix} = \begin{pmatrix} \tilde{\mu}_{x'x'} & 0 \\ 0 & \tilde{\mu}_{y'y'} \end{pmatrix} \begin{pmatrix} \tilde{F}^{mag} \cos\theta \\ -\tilde{F}^{mag} \sin\theta \end{pmatrix} = \begin{pmatrix} \tilde{\mu}_{x'x'} & 0 \\ 0 & \tilde{\mu}_{y'y'} \end{pmatrix} \begin{pmatrix} N\cos\theta \\ -N\sin\theta \end{pmatrix}, \quad (22)$$

where a tilde denotes a variable normalized to that of a single bead as in the previous section. Since we are interested in the velocity in the unprimed system, we apply the rotation matrix,

$$\begin{pmatrix} \cos\theta & -\sin\theta \\ \sin\theta & \cos\theta \end{pmatrix}, \quad (23)$$

to Eq (22) giving

$$\begin{pmatrix} \tilde{U}_x \\ \tilde{U}_y \end{pmatrix} = \begin{pmatrix} \cos\theta & -\sin\theta \\ \sin\theta & \cos\theta \end{pmatrix} \begin{pmatrix} \tilde{\mu}_{x'x'} N \cos\theta \\ -\tilde{\mu}_{y'y'} N \sin\theta \end{pmatrix} = \begin{pmatrix} \tilde{\mu}_{x'x'} N \cos^2\theta + \tilde{\mu}_{y'y'} N \sin^2\theta \\ (\tilde{\mu}_{x'x'} - \tilde{\mu}_{y'y'}) N \cos\theta \sin\theta \end{pmatrix}. \quad (24)$$

**Fig 5. Chain at an angle.** Figure used to derive the analytical expression for the angular dependence of the velocities.

As noted in the previous section in Fig 4, the normalized velocity may be written in the form

$$\begin{pmatrix} \tilde{U}_x \\ \tilde{U}_y \end{pmatrix} = \begin{pmatrix} A + B\cos 2\theta \\ C\sin 2\theta \end{pmatrix} \quad (25)$$

Equating Eq (24) and (25), and solving for *A, B,* and *C*, we obtain

$$\tilde{U}_x(\theta) = \frac{\tilde{\mu}_{x'x'} + \tilde{\mu}_{y'y'}}{2} N + \frac{\tilde{\mu}_{x'x'} - \tilde{\mu}_{y'y'}}{2} N \cos 2\theta \quad (26)$$

and

$$\tilde{U}_y(\theta) = \frac{\tilde{\mu}_{x'x'} - \tilde{\mu}_{y'y'}}{2} N \sin 2\theta. \quad (27)$$

Noting that $\tilde{U}_x = \tilde{\mu}_{x'x'} N$ for $\theta = 0$, and $\tilde{U}_x = \tilde{\mu}_{y'y'} N$ for $\theta = 90°$; and that $\tilde{U}_x = \tilde{U}^{\parallel}$ for $\theta = 0$, and $\tilde{U}_x = \tilde{U}^{\perp}$ for $\theta = 90°$, we may write (including the bead number *N*)

$$\tilde{U}_{N,x}(\theta) = \frac{\tilde{U}_N^{\parallel} + \tilde{U}_N^{\perp}}{2} + \frac{\tilde{U}_N^{\parallel} - \tilde{U}_N^{\perp}}{2} \cos 2\theta \quad (28)$$

and

$$\tilde{U}_{N,y}(\theta) = \frac{\tilde{U}_N^{\parallel} - \tilde{U}_N^{\perp}}{2} \sin 2\theta. \quad (29)$$

Therefore, given the normalized velocity of the chain for both the chain being parallel and perpendicular to the magnetic force, Eqs (28) and (29) give both the $x$ and $y$ components of the normalized velocity for a chain making an arbitrary angle with the magnetic force.

From Eqs (9), (10), and (15) we have

$$\tilde{U}_N^{\parallel} = \frac{3}{2}\left(\ln N + \gamma_N^{\parallel}\right) \tag{30}$$

and

$$\tilde{U}_N^{\perp} = \frac{3}{4}\left(\ln N + \gamma_N^{\perp}\right). \tag{31}$$

These produce excellent fits for $N$ from 2 through 50 as shown in Figs 1 and 2.

Inserting Eqs (30) and (31) into Eqs (28) and (29) we obtain

$$\tilde{U}_{N,x}(\theta) = \frac{3}{8}\left(3\ln N + 2\gamma_N^{\parallel} + \gamma_N^{\perp}\right) + \frac{3}{8}\left(\ln N + 2\gamma_N^{\parallel} - \gamma_N^{\perp}\right)\cos 2\theta \tag{32}$$

and

$$\tilde{U}_{N,y}(\theta) = \frac{3}{8}\left(\ln N + 2\gamma_N^{\parallel} - \gamma_N^{\perp}\right)\sin 2\theta, \tag{33}$$

where $\gamma_N^{\parallel}$ and $\gamma_N^{\perp}$ are given by Eqs (13) and (14). Eqs (32) and (33), with Eqs (13) and (14), provide excellent values for the normalized $x$ and $y$ components of the velocity for a chain of $N$ beads making an arbitrary angle with the magnetic force vector.

## The obliquely falling rod

As noted in the Introduction, previous analyses [26–28] of an obliquely falling rod do not calculate the horizontal and vertical components of the velocity, but calculate only the deflection angle, and then only for the special case of a long thin rod, where the velocity of a vertically falling rod is twice that of a horizontally falling rod. Here we apply the methodology developed here for magnetic bead chains to the obliquely falling rod.

Consider a small cylindrical rod of uniform density falling in a viscous fluid, so that the Reynolds number $\mathrm{Re} = \rho U \ell / \eta \ll 1$ and the system is in the Stokes flow regime. As the gravitational force draws the rod in the direction of the gravitational field, there is a drag force $\mathbf{F}^{drag}$ of the fluid on the rod that is equal and opposite to the effective weight $\mathbf{W}^{eff}$, as illustrated in Fig 6. The effective weight is equal to the gravitational force minus the buoyant force and is assumed to be positive. In addition, the interaction of the fluid with the rod causes the rod to be deflected to the side, producing an additional velocity component $U_x$ in the $+x$ direction, also illustrated in Fig 6.

**Fig 6. Obliquely falling rod.** As the gravitational force draws the rod toward the bottom of the page, there is a drag force that is equal and opposite to the effective weight. The effective weight is equal to the gravitational force minus the buoyant force and is assumed to be positive. In addition, the interaction of the fluid with the rod causes the rod to be deflected toward the right.

Following the same methodology as in the last section, we obtain equations similar to Eqs (28) and (29), that is

$$U_{r,y}(\theta) = \frac{U_r^{\parallel} + U_r^{\perp}}{2} + \frac{U_r^{\parallel} - U_r^{\perp}}{2}\cos 2\theta \tag{34}$$

and

$$U_{r,x}(\theta) = \frac{U_r^{\parallel} - U_r^{\perp}}{2}\sin 2\theta, \tag{35}$$

where the subscript $r$ signifies that we are dealing with a rod, and the tilde was dropped since no normalization is necessary, the gravitational force acting on the rod being independent of the angle $\theta$. Also, $U_r^{\parallel}$ is the speed of a rod falling parallel to its length, that is, a vertically falling rod; and $U_r^{\perp}$ is the speed of a rod falling perpendicular to its length, that is, a horizontally falling rod.

Eqs (34) and (35) give the components of velocity for a uniform cylindrical rod of arbitrary length and arbitrary angle falling through a viscous fluid, in terms of the velocity of a vertically falling rod, the velocity of a horizontally falling rod, and the angle of the rod to the vertical.

Defining $\kappa = U_r^{\parallel}/U_r^{\perp}$ and referring to Fig 6, we find for the deflection angle $\alpha$,

$$\tan\alpha = \frac{U_{r,x}}{U_{r,y}} = \frac{(\kappa-1)\sin 2\theta}{\kappa+1+(\kappa-1)\cos 2\theta} = \frac{(\kappa-1)\tan\theta}{\kappa+\tan^2\theta}, \tag{36}$$

where the expression in terms of tangent was obtained by using double angle formulas for the sine and cosine functions. Note that, since $W^{eff}$ is the same for both vertically and horizontally falling rods, we have $W^{eff} = \zeta^{\parallel}U^{\parallel} = \zeta^{\perp}U^{\perp}$ and can therefore also define $\kappa = \zeta^{\perp}/\zeta^{\parallel}$.

It is interesting to find an expression for $\alpha$ in the limit for a long thin rod. In this limit, $\kappa = 2$ and Eq (36) reduces to

$$\tan\alpha = \frac{\sin 2\theta}{3+\cos 2\theta} = \frac{\tan\theta}{2+\tan^2\theta}, \tag{37}$$

which agrees with the equation for the deflection angle of a long thin obliquely falling rod obtained in Ref [26]. [See Eq (9.32) of that reference. If comparing Eq (9.32) of [26] with Eq (37) of this paper, note that [26] defines the deflection angle as $\theta - \alpha$.]

As noted previously, for a vertically ($\theta = 0°$) or horizontally ($\theta = 90°$) falling rod, there is no deflection of the rod to the side, and thus $\alpha = 0°$. Therefore, as $\theta$ is varied from $0°$ to $90°$, there will be a maximum value of $\alpha$ for some value of $\theta$ between $\theta = 0°$ and $\theta = 90°$. Taking the derivative of Eq (36) with respect $\theta$, and setting it equal to zero then gives the value of $\theta$ for which this maximum occurs. Putting this value of $\theta$ into Eq (36) gives this maximum value of $\alpha$. We thus find for the coordinates of the maxima,

$$(\theta, \alpha)_{max} = \tan^{-1}\left(\sqrt{\kappa}, \frac{(\kappa-1)}{2\sqrt{\kappa}}\right). \tag{38}$$

It may be of interest to plot the deflection angle $\alpha$ as a function of $\theta$ for various values of $\kappa$ from Eq (36). Fig 7 shows the results. Notice that the values of $\kappa$ in Fig 7 range from $\kappa = 1$ to $\kappa = 2$. The value $\kappa = 1$ corresponds to a very short cylindrical rod, for which the ratio of the drag coefficients transverse to and along the rod is equal to 1. As the aspect ratio of the rod is increased, the value of $\kappa$ increases, approaching the asymptotic value of $\kappa = 2$, which is what is referred to as a long thin rod.

**Fig 7. Deflection angle verses angle of rod.** Plot of the deflection angle $\alpha$ as a function of the angle $\theta$ that the rod makes with the vertical for several values of $\kappa$ for an obliquely falling rod as given by Eq (36). In addition, the figure gives the coordinates for the maxima of the plots using Eq (38), as well as the coordinates for the maxima for a continuous range of values for $\kappa$ from $\kappa = 1$ to $\kappa = 2$ using Eq (38), as indicated by the dotted line.

In addition to the plots of $\alpha(\theta)$ for various values of $\kappa$, Fig 7 also gives the coordinates for the maxima of the curves using Eq (38), as well as a plot of the coordinates for the maxima for a continuous range of values for $\kappa$ from $\kappa = 1$ to $\kappa = 2$ using Eq (38), as indicated by the dotted line. Notice that for $\kappa = 1$ the deflection angle $\alpha$ is zero for all values of the rod angle $\theta$, since the velocities are the same for both the vertical and horizontal orientations of the rod. Also, the maximum value for $\kappa = 2$ corresponds to a long thin rod.

Since the analyses in Ref [26–28] for a long thin obliquely falling rod use a geometric approach, it may be useful to derive Eq (36) using such an approach here. Fig 8 shows the rod with both the drag force and the velocity resolved into components parallel and perpendicular to the rod. Referring to Fig 8, we find the following equations:

$$F_{y'}^{drag} = -\zeta^{\parallel} U_{y'} \tag{39}$$

and

$$F_{x'}^{drag} = -\zeta^{\perp} U_{x'}, \tag{40}$$

where $\zeta^{\parallel}$ and $\zeta^{\perp}$ are the drag coefficients for a rod moving parallel and perpendicular to the length of the rod, respectively. Noting that $\kappa = \zeta^{\perp}/\zeta^{\parallel}$, dividing Eq (40) by Eq (39), and referring to Fig 8 we find

$$\tan\theta = \kappa \tan(\theta - \alpha). \tag{41}$$

Using $\tan(\theta - \alpha) = (\tan\theta - \tan\alpha)/(1 + \tan\theta \tan\alpha)$ and solving for $\tan\alpha$ we find for the deflection angle $\alpha$,

$$\tan\alpha = \frac{(\kappa - 1)\tan\theta}{\kappa + \tan^2\theta}, \tag{42}$$

which is the same as that given by Eq (36). As expected, we see that the deflection of the rod to the right is the result of the drag coefficient for motion along the rod being less than that for motion transverse to the rod.

**Fig 8. Geometric derivation of the deflection angle.** Figure used for the geometric derivation of the deflection angle. Both the drag force and the velocity are resolved into components parallel to and perpendicular to the length of the rod. Because the drag coefficient for motion along the rod is less than that for motion transverse to the rod, the rod is deflected toward the right.

## Conclusions

In this paper we used the public-domain computer program HYDRO++ to calculate the mobility matrix for a chain of $N$ beads making an arbitrary angle with the magnetic force vector. We found that, as the chain is drawn in the direction of the magnetic force, the chain is also deflected to the side due the interaction of the chain with the fluid, causing a component of velocity perpendicular to the magnetic force vector. This sideways motion results from the presence of off-diagonal elements of the mobility matrix. We derive analytical expressions for the components of the normalized velocity both parallel and perpendicular to the magnetic force in terms of the normalized velocity of a chain that is parallel to the magnetic force, the normalized velocity of a chain that is perpendicular to the magnetic force, and the angle that the chain makes with the magnetic force.

We also fit three-parameter functions to the normalized drag coefficients calculated from HYDRO++ solutions for the chain being both parallel and perpendicular to the magnetic force. Combining these fits with the analytical solutions for a chain that makes an arbitrary angle with the magnetic force, we give expressions for the components of the normalized velocity both parallel and perpendicular to the magnetic force for a chain of $N$ beads making an arbitrary angle with the magnetic force vector.

Finally, we apply the methodology used for the bead chains to the case of an obliquely falling rod in a viscous fluid in a gravitational field. We derive more general expressions than those found in the literature [26–28] for the horizontal and vertical components of velocity in terms of the velocity of a vertically falling rod, the velocity of a horizontally falling rod, and the angle that the rod makes to the vertical.

## Acknowledgements

We are grateful to Professors Jesse Berezovsky, Michael Hinczewski, and Michael Martens for general discussions about magnetic particle physics.

# Figures

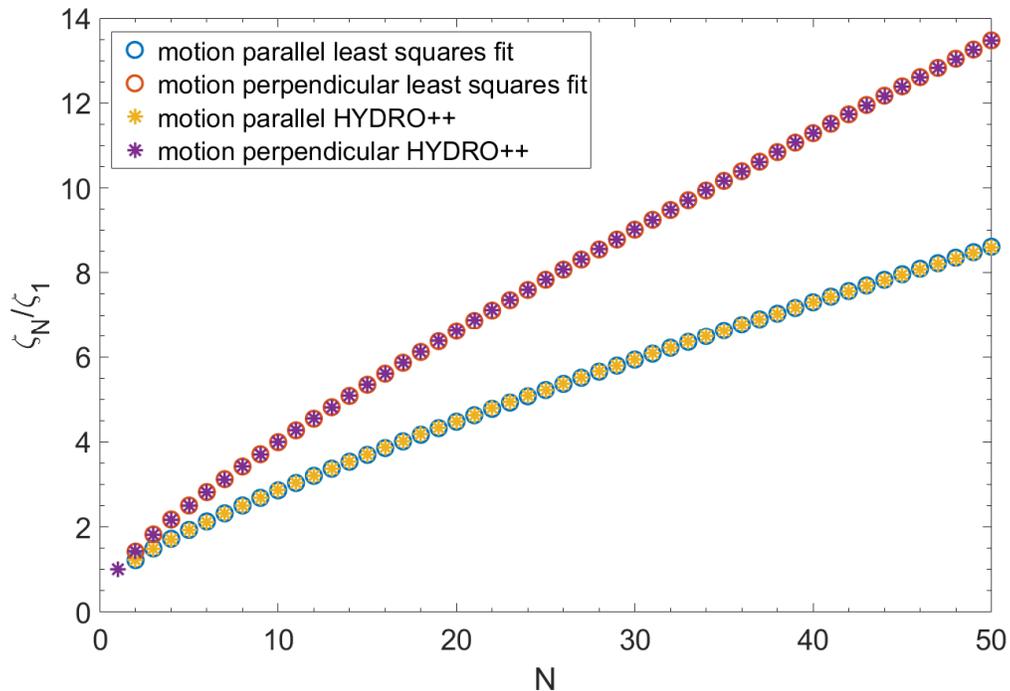

**Fig 1. Normalized drag coefficients.** Plot of the normalized drag coefficients for a chain of $N$ beads for motion both parallel and perpendicular to the length of the chain from the numerical solution using HYDRO++ and from Eq. (9), (10), (13), and (14), which resulted from a least squares fit to the numerical solution of HYDRO++. Agreement is seen to be excellent.

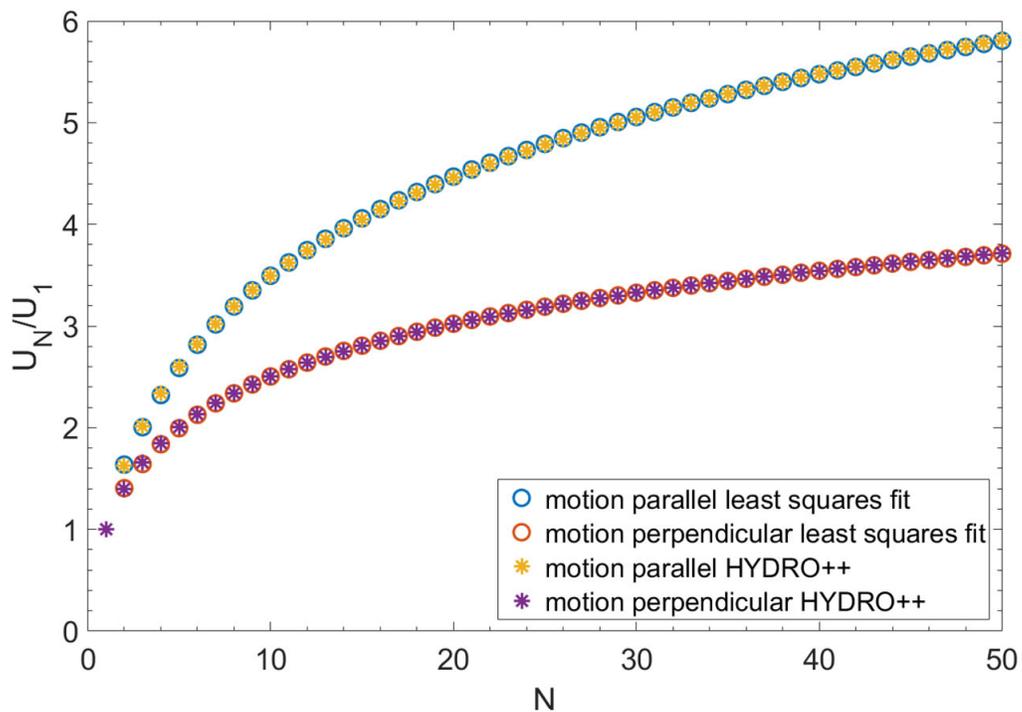

**Fig 2. Normalized translational velocity.** Plot of the normalized velocity for a chain of *N* beads for motion both parallel and perpendicular to the length of the chain from the numerical solution using HYDRO++ and from Eq. (9), (10), (13), (14), and (15) which resulted from a least squares fit to the numerical solution of HYDRO++.

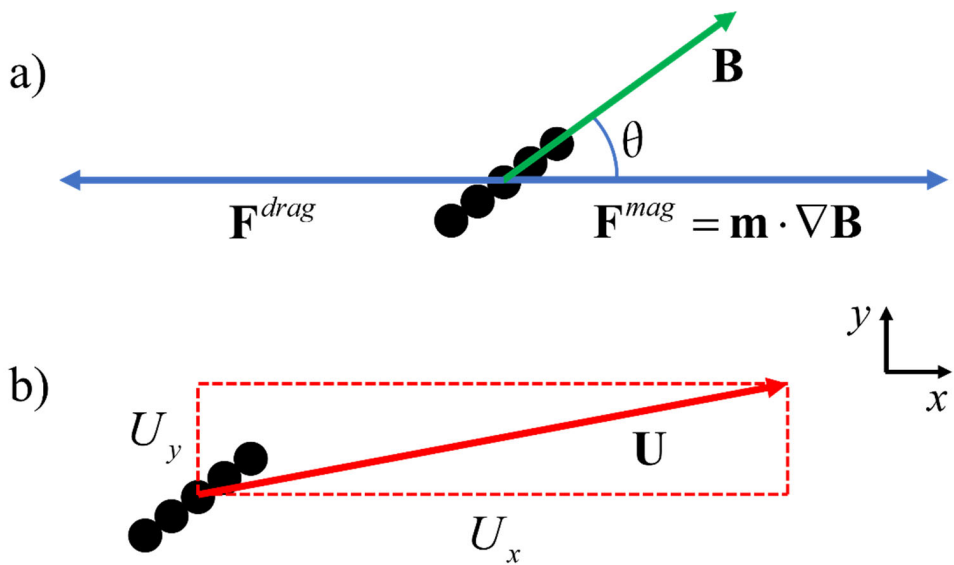

**Fig 3. Chain at an angle.** a) As the magnetic force draws the chain to the right in the +*x* direction, the interaction of the fluid with the chain produces a drag force in the -*x* direction. b) In addition, the interaction of the fluid with the chain causes the chain to be deflected to the side, producing an additional velocity component in the +*y* direction, which is related to the off-diagonal elements of the mobility matrix.

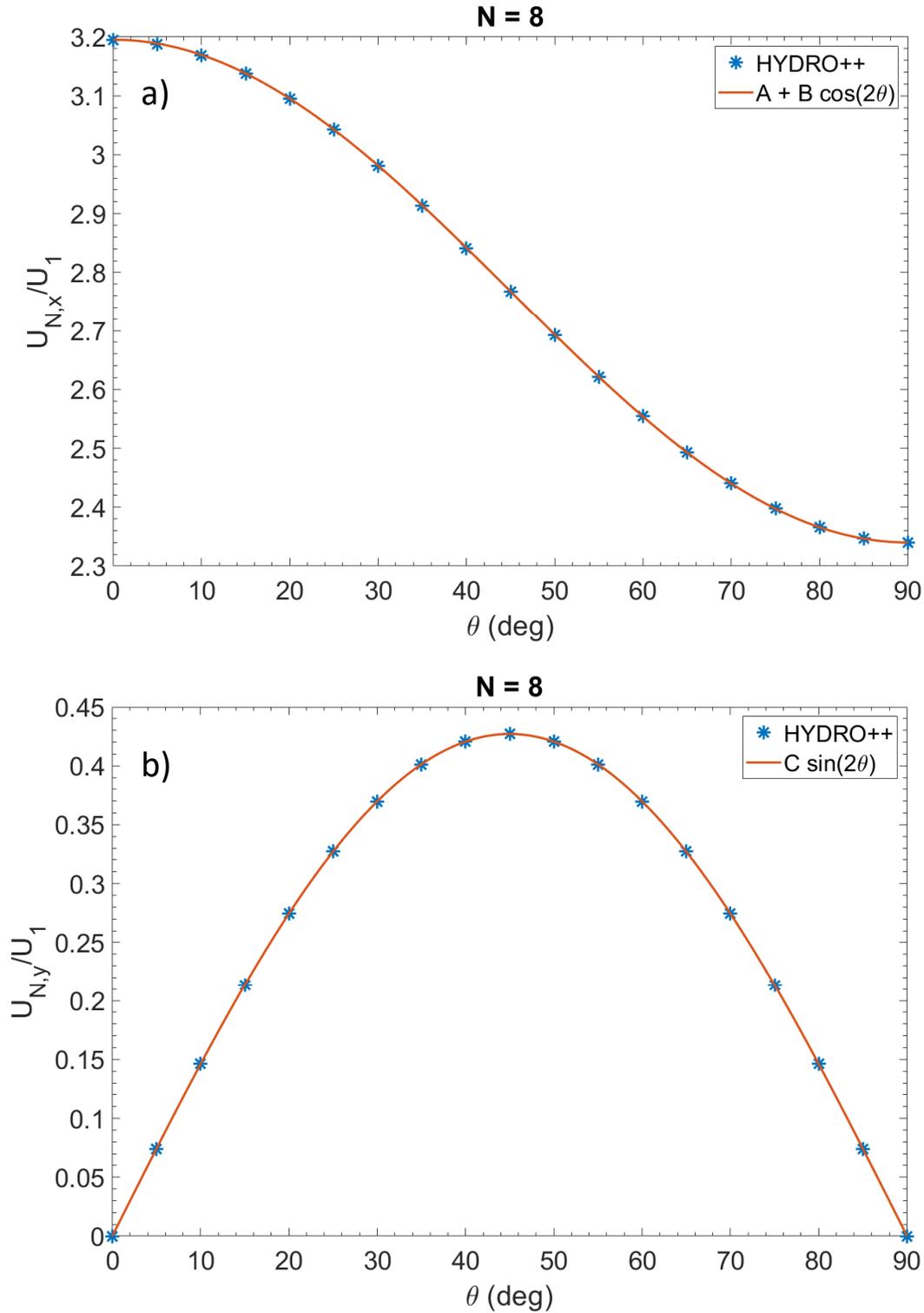

**Fig 4. Normalized velocities in the x and y directions for a chain of 8 beads for a range of angles.** a) $x$ component of the velocity. b) $y$ component of the velocity. It is seen that the cosine and sine functions fit precisely with the HYDRO++ solutions. The coefficients $A$, $B$, and $C$ are determined from $A = [\tilde{U}_x(0°) + \tilde{U}_x(90°)]/2 = 2.76758$, $B = [\tilde{U}_x(0°) - \tilde{U}_x(90°)]/2 = 0.42757$, and $C = \tilde{U}_y(45°) = 0.42701$, respectively.

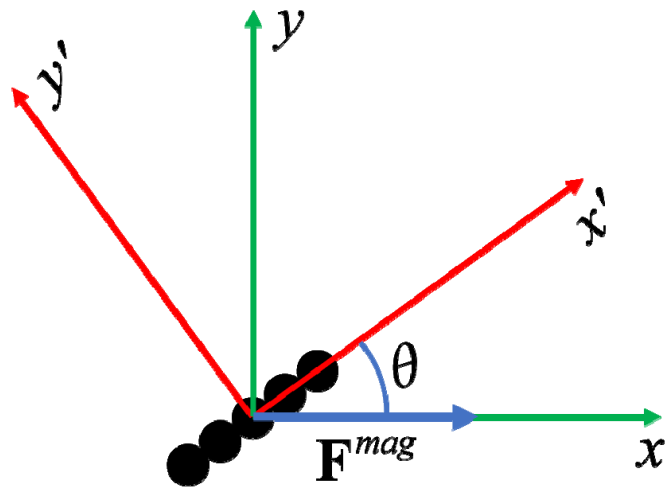

**Fig 5. Chain at an angle.** Figure used to derive the analytical expression for the angular dependence of the velocities.

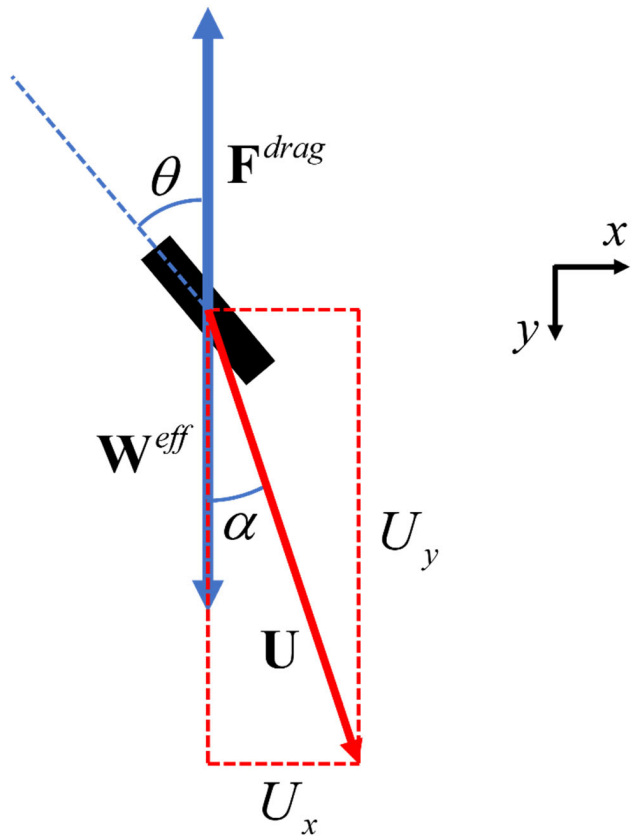

**Fig 6. Obliquely falling rod.** As the gravitational force draws the rod toward the bottom of the page, there is a drag force that is equal and opposite to the effective weight. The effective weight is equal to the gravitational force minus the buoyant force and is assumed to be positive. In addition, the interaction of the fluid with the rod causes the rod to be deflected toward the right.

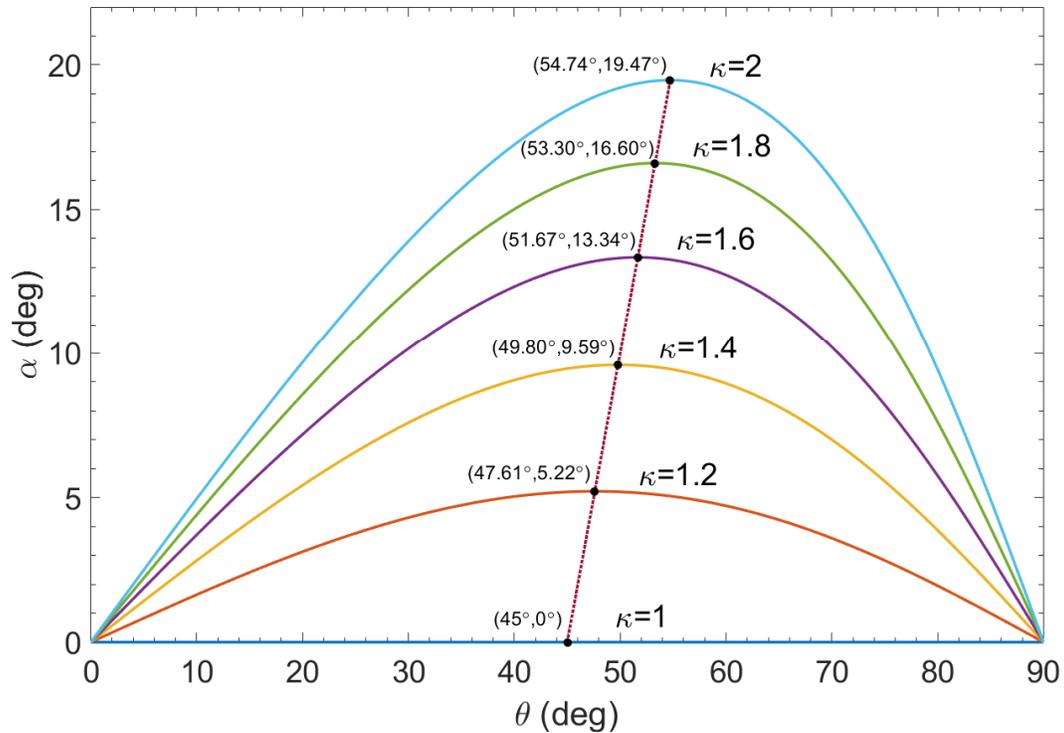

**Fig 7. Deflection angle verses angle of rod.** Plot of the deflection angle $\alpha$ as a function of the angle $\theta$ that the rod makes with the vertical for several values of $\kappa$ for an obliquely falling rod as given by Eq (36). In addition, the figure gives the coordinates for the maxima of the plots using Eq (38), as well as the coordinates for the maxima for a continuous range of values for $\kappa$ from $\kappa=1$ to $\kappa=2$ using Eq (38), as indicated by the dotted line.

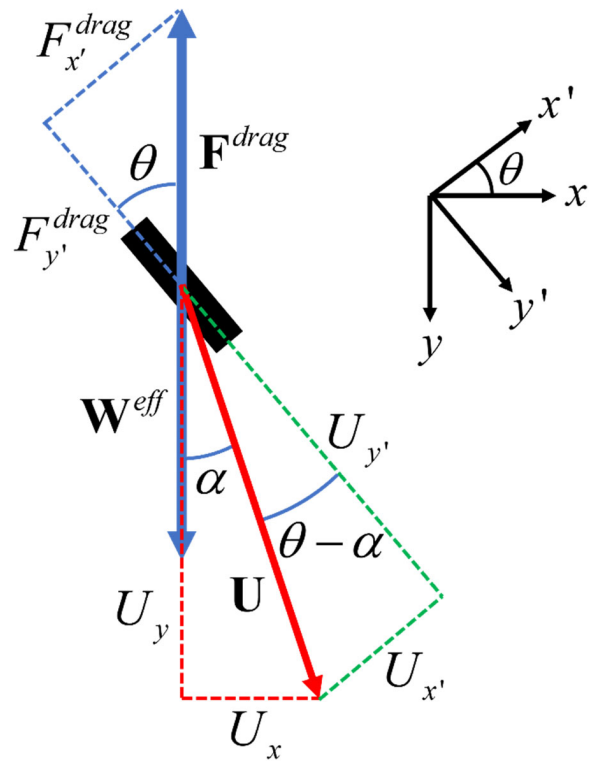

**Fig 8. Geometric derivation of the deflection angle.** Figure used for the geometric derivation of the deflection angle. Both the drag force and the velocity are resolved into components parallel to and perpendicular to the length of the rod. Because the drag coefficient for motion along the rod is less than that for motion transverse to the rod, the rod is deflected toward the right.